\documentclass[prc,preprint]{revtex4}
\usepackage{graphicx,float}
\usepackage{amssymb,amsmath}
\usepackage{graphicx}
\usepackage{epstopdf}
\usepackage{bbding}
\usepackage{pifont}
\usepackage{wasysym}

\begin{document}

\title{\bf Structure of hot strange quark stars: an NJL model approach at finite temperature}

\author{{ G. H. Bordbar$^{1,2,3}$
\footnote{Corresponding author. E-mail:
ghbordbar@shirazu.ac.ir}}, { R. Hosseini $^{1}$} and { F. Kayanikhoo  $^{4}$} and {A. Poostforush $^{1}$}}
\affiliation{$^1$ Department of Physics and Biruni Observatory,
Shiraz University, Shiraz 71454, Iran\footnote{Permanent address}\\
$^2$Research Institute for Astronomy and Astrophysics of Maragha,
PO Box 55134-441, Maragha, Iran\\
$^3$ Department of Physics and Astronomy, University of Waterloo,
200 University Avenue West, Waterloo, Ontario,
N2L3G1,
Canada\\
$^4$ Department of Physics, University of Birjand, Birjand, Iran
}
%
\begin{abstract}

In this paper, we investigated the thermodynamic properties of strange quark matter using Nambu-Jona-Lasinio (NJL) model at finite temperatures
where we considered the dynamical mass as the effective interaction between quarks.
 By considering the pressure
of strange quark matter (SQM) at finite temperatures, we showed that the equation of state of this system gets stiffer by increasing temperature.
In addition, we investigated the energy conditions and stability of the equation of state and showed that the equation of state of SQM satisfy the conditions of stability.
Finally, we computed the structure properties of hot strange quark stars (SQS) including the gravitational mass, radius, Schwarzschild radius, average density, compactness and gravitational redshift.
Our calculations showed that in this model, the maximum mass and radius of SQS increase by increasing temperature.
Furthermore it was shown that the average density of SQS is greater than the normal nuclear density, and it is an increasing function of temperature.
We also discussed the temperature dependence of the maximum gravitational mass calculated from different methods.
 \end{abstract}

\maketitle

\section{Introduction}
By the mid-1970s, physicists realized that hadrons are made up of new particles later called quarks
with a model first proposed by Gell-Mann and Zwieg \cite{Gell-Mann1964, Zweig1964}.
Baryons at high enough densities (≈$10^{15}$ $gr/cm^{3}$) overlap and dissolve to their components, quarks.

The concept of strange quark matter (SQM), dates back to the works of Jaffe \cite{Jaffe1977}, Chin and Kerman \cite{Chin1979}. SQM contains the light quarks (up, down and strange).
 In 1984, Witten \cite{Witten1984} proposed that SQM might be absolutely stable, and might be the true ground state of baryonic matter.

Strange quark stars (SQS) are compact objects  interesting for astrophysicists and physicists as the SQS is a great laboratory to study the properties of SQM due to
the density of  about $10^{15}$ $g/cm^{3}$. The composition of SQS was first proposed by Itoh \cite{Itoh1970} with the formulation of quantum chromodynamics (QCD).
In 1971, Bodmer \cite{Bodmer1971} discussed the possibility of forming a quark star after the collapse of a massive star,
later the concept of SQS  was also mentioned by Witten \cite{Witten1984}.
%

The collapse of a massive star could lead to the formation of a pure SQS by type $IIa$ supernova (SNII) \cite{Sato1987, Suzuki1987, Hatsuda1987}.
 Also, a hybrid star which is a neutron star with a core consisting SQM, can be formed after neutron star, if the density of the core is high enough.
The recent Chandra observations indicate that objects \emph{RXJ185635-3754} and \emph{3C58} may be SQSs \cite{Prakash2003}, as well as candidate for SQS is the object \emph{SWIFTJ1749.4-2807} \cite{Yu2010}.
Actually, a SQS or a hybrid star is denser than a neutron star. In other words, the mass of SQS is near that of a neutron star but with a smaller radius.

There are two main frameworks usually used to investigate the thermodynamic properties of SQM, Nambu-Jona-Lasinio (NJL) model \cite{Carroll2009} and MIT bag model \cite{Chodos1974, Alford2005},
where theoretical foundations of both is QCD \cite{Freedman1977}.
In recent years, we have investigated the thermodynamic properties of SQM and structure of SQS under different conditions using these frameworks.
We have computed the structural properties of SQS at zero  and finite temperatures,
as well as the structure of a magnetized SQS using the MIT bag model with the fixed and density dependent bag constants at zero and finite temperatures in the presence and absence of magnetic fields
\cite{Bordbar2011vol11, Bordbar2011vol54, Bordbar2012vol12, Bordbar2013, Bordbar2014}.
 We have also computed the maximum gravitational mass and other structural properties of a neutron star with a quark core at zero \cite{Bordbar2006}
 and finite temperatures \cite{Yazdizadeh2011}.

In our previous work, we have studied the effect of
 dynamical quark mass in the calculation of SQS structure using MIT bag model and NJL model at zero temperature \cite{Bordbar2012vol12540}.
In current paper, we extend  NJL model for finite temperatures to survey the thermodynamic properties of a hot SQS.
Furthermore, we show that the equation of state of SQM calculated according to NJL model satisfies the stability and energy conditions.
We investigate the structure properties of SQS by calculating the structure parameters
 (mass, radius, Schowarzschild radius, average density, compactness and gravitational redshift) in the last section.
%

\section{Calculation of energy and equation of state of hot SQM using NJL model}
\label{II}
\subsection{Nambu-Jona-Lasinio (NJL) model at finite temperatures}
The NJL model is named after Nambu, Jona and Lasinio, who for the first time offered a theory about the
dynamical model of elementary particles based on analogy with super conductivity in $1961$ \cite{Nambu1961}.
The NJL model is an effective lagrangian of relativistic fermions interacting through local fermion-fermion coupling.
This model is a suitable approximation of QCD in the low energy and long wavelength limits \cite{Klevansky1992, Vogl1991},
appropriate for the bound states of many-body systems \cite{Buballa2005} and EOS of compact stars \cite{Schertler1999}.
At high temperatures and densities, interaction leads to spontaneously breaking of chiral symmetry. In NJL model, symmetry breaking is characterized by quarks dynamical mass \cite{Peng1999, Shao2011}.

Dynamical mass of quarks is calculated via,
\begin{equation}\label{01}
M_{i}=m_{0}^{i}-4G<\overline{q_{i}}q_{i}>+2K<\overline{q_{j}}q_{j}><\overline{q_{k}}q_{k}>,
\end{equation}
where $M_{i}$ is dynamical mass of quark $i$, $m^{i}_{0}$ is mass of free quark $i$,
$G$ and $K$ are the coupling constants, and $<\overline{q_{i}}q_{i}>$ shows the condensation of quark-antiquark which is calculated as follows
\cite{Pennington2005, Buballa2005},
\begin{equation}\label{02}
<\overline{q_{i}}q_{i}>=- \frac{3}{\pi^{2}(\hbar c)^{3}}\int_{p_{f}^{i}}^{\lambda}\frac{M_{i}}{\sqrt{M_{i}^{2} c^{2}+p^{2} c^{2}}}f_{i}(p)p^{2}dp.
\end{equation}
In the above equation, $\lambda$  in the upper limit of integral is the cut-off value, $p$ is momentum of quark,
$p_{f}^{i}$ is the Fermi momentum of each quark and
\begin{equation}\label{03}
f_{i}(p)=\frac{1}{e^{\beta{(\epsilon_{i}(p)-\mu_{i})}}+1},
\end{equation}
where $\mu_{i}$ and $\epsilon_{i}$ are the chemical potential and single particle energy of quark $i$, respectively, and $\beta$ $=$ $1/k_{B}T$ ($k_{B}$ is the Boltzmann constant).
We calculate the chemical potential, $\mu_{i}$ by solving the Fermi-Dirac equation numerically.
The NJL model is a renormalizable model,
so we should choose a method to find the physical values.
In the present paper, we use an ultra-violent cut-off that indicates restoring of chiral symmetry breaking, $\lambda=602.3$ $MeV$ \cite{Raha2000, Buballa2005}.

\subsection{Energy and EOS of hot SQM}

The total energy density of SQM is defined as the sum of the kinetic energy of free quarks, $\varepsilon_{i}$, and
the potential energy of our system, $B_{eff}$, which is called the effective bag constant,
\begin{equation}\label{04}
\varepsilon_{tot}=\sum_{i=u,d,s} \varepsilon_{i} +B_{eff},
\end{equation}
where the kinetic energy of quark $i$ ($\varepsilon_{i}$) is calculated using the following constraint,
\begin{equation}\label{05}
\varepsilon_{i}=-\frac{3}{\pi^{2}(\hbar c)^{3}}\int_{0}^{\lambda} \sqrt{M_{i}^{2} c^{2}+p^{2} c^{2}}f_{i}(p)p^{2}dp.
\end{equation}
The effective bag constant is calculated by the following relation,
\begin{equation}\label{06}
B_{eff}=B_{0}+B_{tot},
\end{equation}
where,
\begin{equation}\label{07}
B_{tot}=\sum_{i}B_{i}+\frac{1}{(\hbar c)^{3}}4K<\overline{u}u><\overline{d}d><\overline{s}s>,
\end{equation}
and,
\begin{equation}\label{08}
B_{i}=\frac{3}{\pi^{2}(\hbar c)^{3}}\int_{0}^{\lambda}\left[\sqrt{M_{i}^{2} c^{2}+p^{2} c^{2}}-\sqrt{m_{i}^{2} c^{2}+p^{2} c^{2}}-2G<\overline{q_{i}}q_{i}>^{2}\right]p^{2}dp.
\end{equation}
We need the Helmholtz free energy to calculate the equation of state (EOS) of the system,
\begin{equation}\label{09}
{\cal F}_{tot}=\varepsilon_{tot}-TS_{tot},
\end{equation}
where $S_{tot}$ is the total entropy of system,
\begin{equation}\label{10}
S_{tot}=\sum_{i=u,d,s}s_{i},
\end{equation}
and $ s_{i}$ is entropy of quark $i$,
\begin{equation}\label{11}
s_{i}=-\frac{3}{\pi^{2}(\hbar c)^{3}}\int_{0}^{\lambda}\left\{f_{i}(p)\ln{f_{i}(p)}+[1-f_{i}(p)]\ln{[1-f_{i}(p)]}\right\}p^{2}dp.
\end{equation}
To calculate EOS of our system, we use the following relation,
\begin{equation}\label{12}
P(n,T)=\sum_{i}n_{i}\frac{d{\cal F}_{tot}}{dn_{i}}-{\cal F}_{tot},
\end{equation}
where $n_{i}$  is the number density of quark $i$.

\subsection{Results of thermodynamic properties of hot SQM}
\begin{figure}
\centering
\includegraphics[width=10cm, height=7cm]{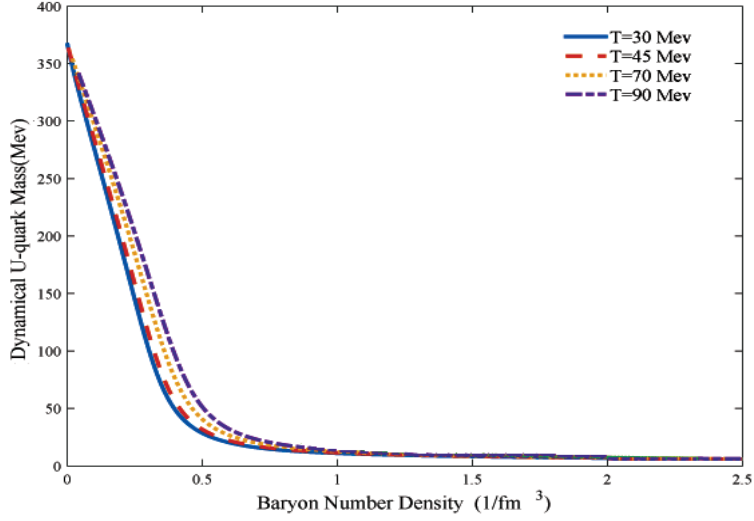}
\caption{Dynamical mass of $up$ quark versus baryon number density at different temperature.} \label{fig1}
\end{figure}
\begin{figure}
\centering
\includegraphics[width=10cm, height=7cm]{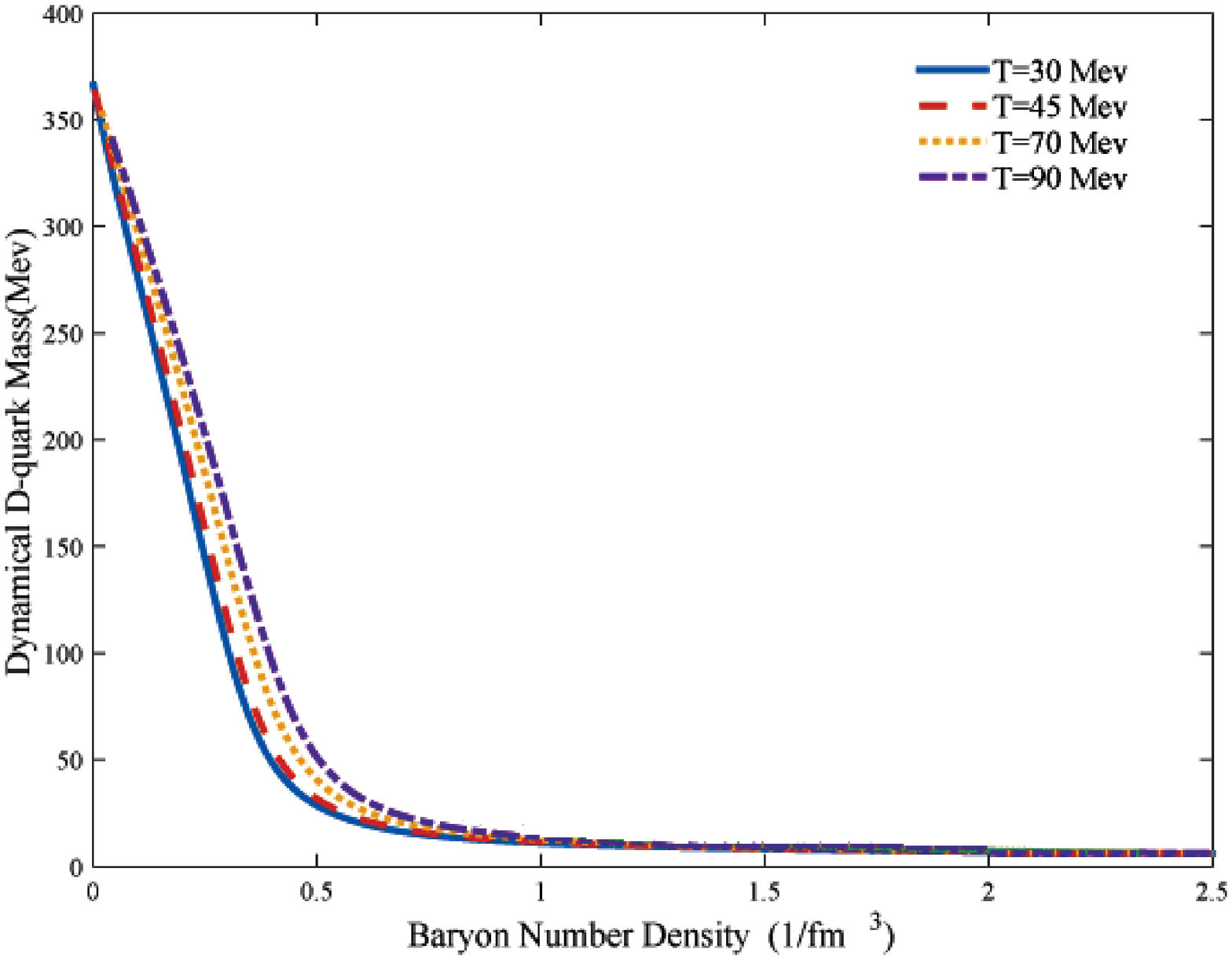}
\caption{Dynamical mass of $down$ quark versus baryon number density at different temperature.} \label{fig2}
\end{figure}
\begin{figure}
\centering
\includegraphics[width=10cm, height=7cm]{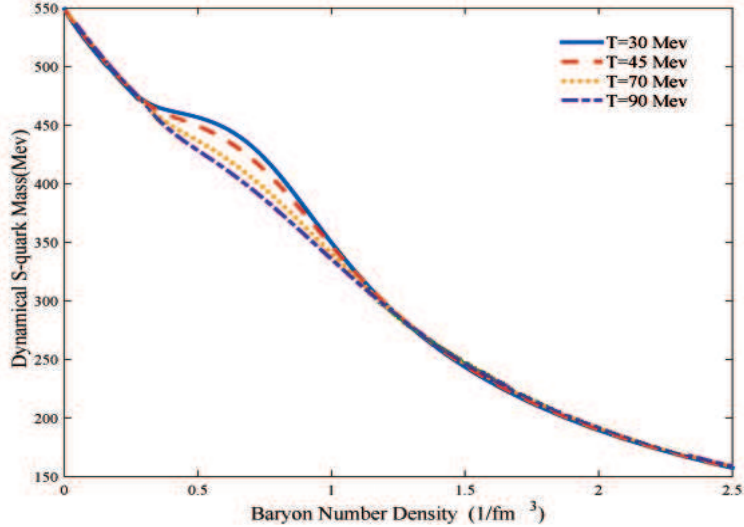}
\caption{Dynamical mass of $strange$ quarks versus baryon number density at different temperature.} \label{fig3}
\end{figure}
%
In Figs.\ref{fig1}, \ref{fig2} and \ref{fig3}, we have presented the dynamical mass of the
up, down and strange quarks versus baryonic number density, respectively.
We compare our results at different temperatures. As the plots show, the
dynamical mass of each quark tend to the inertial mass
($m_{s}=140.7$ $MeV$ and $m_{u}=m_{d}=5.5$ $MeV$) by increasing the baryonic density.
Also, we can see for $u$ and $d$ quarks, the dynamical mass increases by increasing the temperature.
These results hold for strange quarks as well, except for density of $0.5-1\ fm^{-3}$.
Our results are also consistent with the previous ones \cite{Bordbar2012vol12540} and the results of Ruster et al. \cite{Ruster2006}.

\begin{figure}
\centering
\includegraphics[width=10cm, height=7cm]{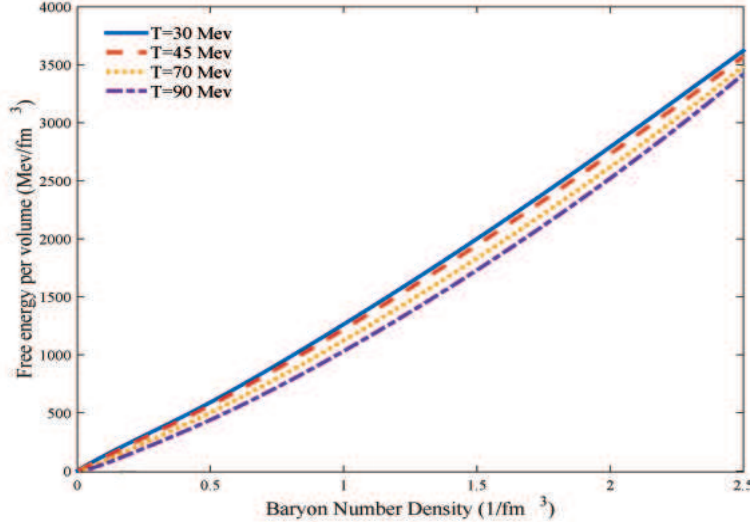}
\caption{The total free energy per volume of SQM as a function of the baryonic
density at different temperatures.} \label{fig4}
\end{figure}

We have shown the total free energy per volume of hot SQM
as a function of the baryonic density in Fig. \ref{fig4}. By increasing the baryonic number density, the free energy increases.
Also it is seen that the free energy decreases with increasing temperature.
%

\begin{figure}
\centering
\includegraphics[width=10cm, height=7cm]{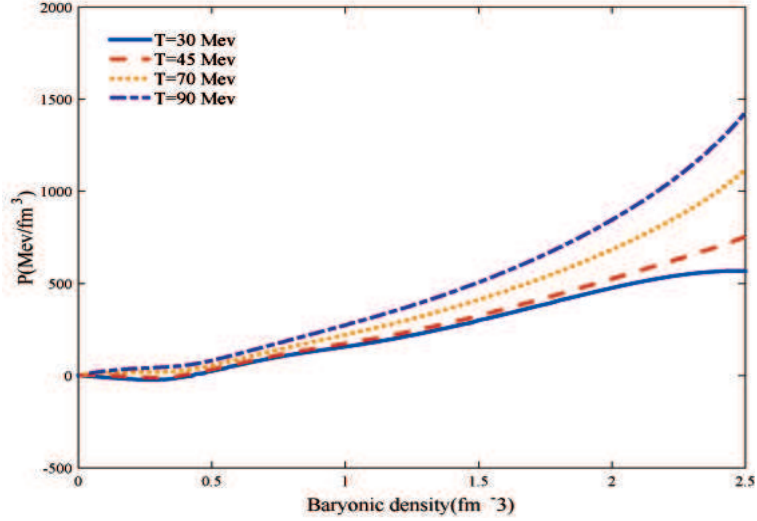}
\caption{The pressure of SQM at different temperatures.} \label{fig5}
\end{figure}

The pressure of hot SQM at different temperatures has been plotted in Fig. \ref{fig5}.
%
This figure shows that the pressure of SQM increases
by increasing the density. We can also see that  the pressure increases by increasing the temperature.
These results indicate that the equation of state of SQM becomes stiffer by increasing the temperature.
In other words, the compressibility of the degenerate gas decreases by increasing temperature, therefore, the EOS becomes stiffer.
\begin{figure}
\centering
\includegraphics[width=10cm, height=7cm]{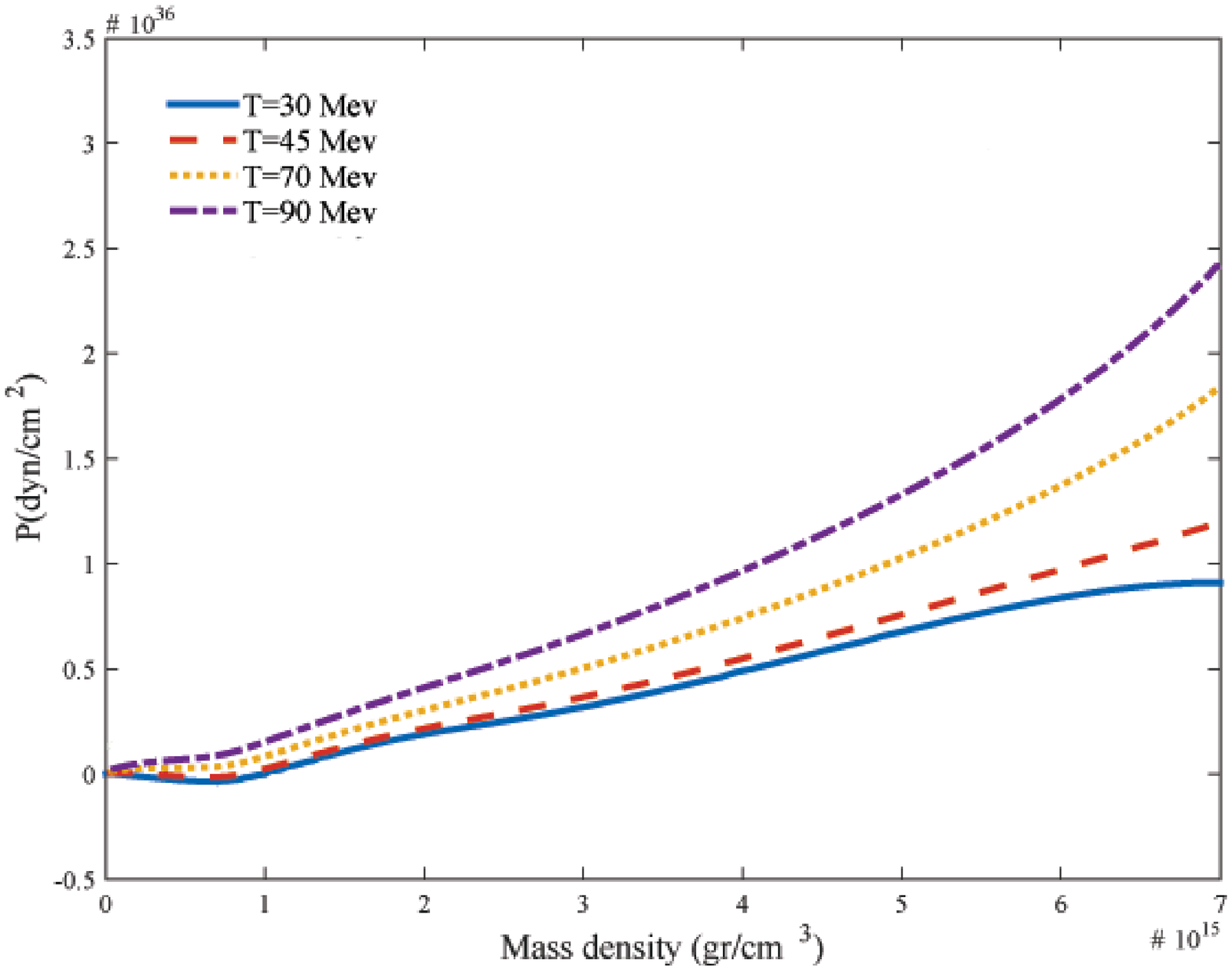}
\caption{The pressure of SQM versus mass density at different temperatures.} \label{fig06}
\end{figure}
%
In Fig. \ref{fig06} we have plotted the pressure of SQM versus mass density at different temperatures.
Our results show that the pressure increases by increasing mass density. Also, it is shown that the central pressure increases as a function of temperature.

{\bf Here, we show that in the considered version of the equation of state, the Bodmer-Witten hypothesis holds true.
According to this hypothesis, the energy per particle of SQM should be lower than
that of $^{56}Fe$ which is $930.4$$MeV$, so SQM is more stable than the nuclear matter~\cite{Bodmer1971, Witten1984}.
To investigate this condition, we have investigated the energy per particle behavior at different temperatures ($T$).
We have found that the minimum point of energy per particle versus baryon density which is corresponding to the zero pressure is equal to $408.77$$MeV$ at
$T=30$$MeV$ and is equal to $928.55$$MeV$ at  $T=90$$MeV$ which ensures the stability of SQM.
We also study energy and stability conditions in next parts.}

\subsection{Energy conditions}
There are four different energy conditions that we study in this work;\\
\\a) Null energy condition (NEC) $\longrightarrow$ $P_{c}$ + $\rho_{c} c^{2}$$ \geq$0,\\
b) Weak energy condition (WEC)  $\longrightarrow$ $P_{c}$ + $\rho_{c} c^{2}$$ \geq$0 and  $\rho_{c}$$ \geq$ 0,\\
c) Strong energy condition (SEC) $\longrightarrow$ $P_{c}$ + $\rho_{c} c^{2}$$ \geq$ 0 and $3P_{c}$ + $\rho_{c} c^{2}$$ \geq$ 0,\\
d) Dominate energy condition (DEC) $\longrightarrow$ $ \rho_{c} c^{2}$$ \geq $$\mid$ $P_{c}$ $\mid$,\\
\\
where $\rho_{c}$ and $P_{c}$ are mass density and pressure at the center of SQS ($r=0$).
Results shown in Table \ref{T1} at different temperatures correspond to  Fig. \ref{fig06} and the above four conditions.
It is clear that all energy conditions are satisfied regarding the equation of state we calculated for SQM.
%
\begin{table}[h]
\begin{center}
  \caption[]{Energy conditions of SQS
at different temperatures.}\label{T1}
  \begin{tabular}{clclclclclclcl}
  \hline\noalign{\smallskip}
$T (MeV)$ && $\rho_{c} (10^{15}$$\frac{ g} {cm^{3}}$) & $P_{c} (10^{15}$$\frac{ g} {cm^{3}}$) && $NEC$ && $WEC$ && $SEC$ && $DEC$\\
 \hline\noalign{\smallskip}
 $30$ && 7 & 0.969 && $\checkmark$ && $\checkmark$ && $\checkmark$ && $\checkmark$ \\
 $45$ && 7 & 1.315 && $\checkmark$ && $\checkmark$ && $\checkmark$ && $\checkmark$ \\
 $70$ && 7 & 2.007 && $\checkmark$ && $\checkmark$ && $\checkmark$ && $\checkmark$  \\
 $90$ && 7 & 2.630 && $\checkmark$ && $\checkmark$ && $\checkmark$ && $\checkmark$  \\
 $150$ && 6.5 & 4.153 && $\checkmark$ && $\checkmark$ && $\checkmark$ && $\checkmark$ \\

  \noalign{\smallskip}\hline
  \end{tabular}
\end{center}
\end{table}
%
\subsection{Stability of equation of state}
To verify the stability of EOS of SQM we use the extreme condition of sound velocity. The sound velocity is calculated by $v_{s}=\sqrt{dP/d\rho}$.
It is clear that to have a physical model, the sound velocity must satisfy the condition of $0$ $ \leq$$v_{s}^{2}$$\leq$ $c^{2}$.
Here we have found that for all relevant densities and temperatures, the above condition is obeyed by the velocity of sound.
This indicates that the stability of our EOS is confirmed for all temperatures and densities except densities less than $0.7 \times 10^{15}$$gr/cm^{3}$ at temperature of $30$$MeV$.
It is clear that the strange quark matter can be created at high enough temperature and density \cite{Nakazato2008, Nakazato2010, Nakazato2013}.
%
%

%
\section{Structure properties of strange quark star}
\label{struc}
The structure of stars is usually determined by their mass and radius, although there are some other parameters, such as schwarzschild radius, average density, compactness and gravitational redshift, which we investigate.

\subsection{Mass and radius of SQS}\label{secA}
Since quark stars are relativistic objects we should use the relativistic equation of hydrostatic equilibrium for these systems,
\begin{equation}\label{13}
\frac{dP}{dr}=-\frac{G\left[\varepsilon(r)+\frac{P(r)}{c^{2}}\right]\left[m(r)+\frac{4\pi
r^{3}P(r)}
{c^{2}}\right]}{r^{2}\left[1-\frac{2Gm(r)}{rc^{2}}\right]},
\end{equation}
\begin{equation}\label{14}
\frac{dm}{dr}=4\pi r^{2}\varepsilon(r).
\end{equation}
These equations are known as Tolman-Oppenheimer-Volkov equations (TOV) \cite{Oppenheimer1939}.
Using the equation of state which was obtained in the previous
section and the boundery conditions ($P(r=0)=P_{c}$, $P(r=R)=0$, $m(r=0)=0$ and $m(r=R)=M_{max}$)
we integrate the TOV equations to compute the structure of strange quark stars
(SQS).
\begin{figure}
\centering
\includegraphics[width=10cm, height=7cm]{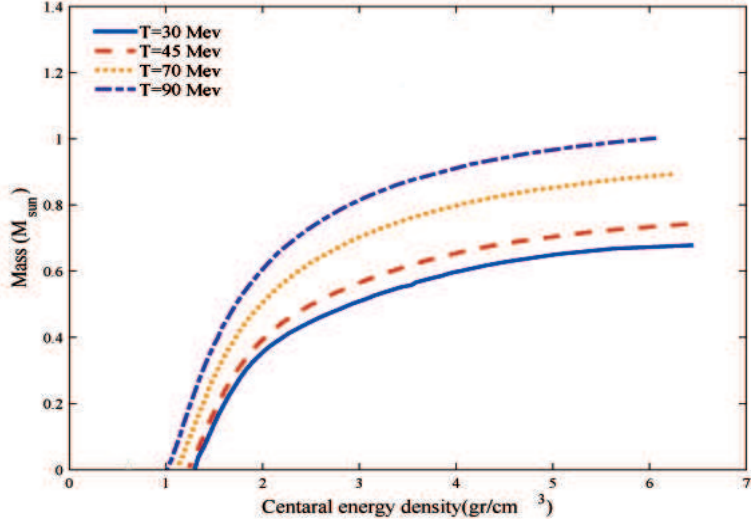}
\caption{The gravitational mass of SQS versus energy density at different temperatures.} \label{fig6}
\end{figure}

\begin{figure}
\centering
\includegraphics[width=10cm, height=7cm]{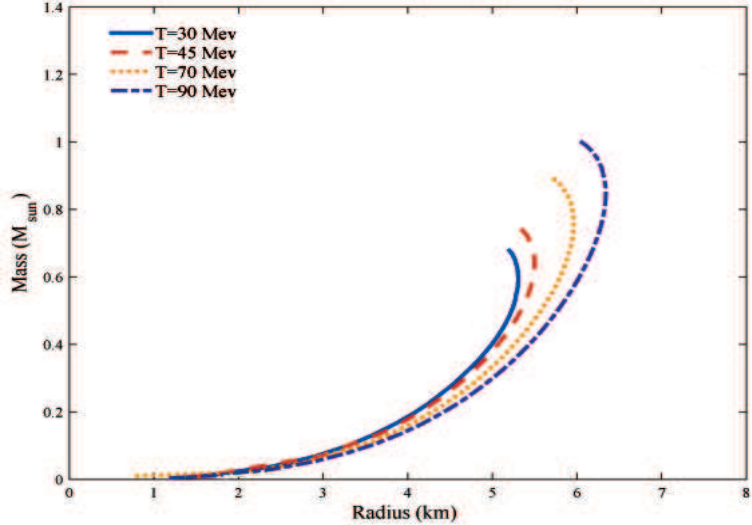}
\caption{The gravitational mass of SQS as a function of the radius at different temperatures.} \label{fig7}
\end{figure}

In Fig. \ref{fig6}, we have plotted the gravitational mass of strange quark star
(SQS) versus energy density at different temperatures.
We can see that for all temperatures, the gravitational mass
increases rapidly by increasing the energy
density, and finally reaches a limiting value (maximum gravitational mass).
The maximum gravitational mass for different temperatures has been given in Table \ref{T2}.
Our results show that this maximum mass increases by increasing the temperature.
We have shown the gravitational mass of  SQS as a function of the radius (M-R relation) at different
temperatures in Fig. \ref{fig7}.
%
This figure shows that by increasing the gravitational mass, till the maximum mass is reached, the radius increases.
We can see that the increasing rate of gravitational mass versus radius
increases by increasing temperature. The radius of SQS corresponding to the maximum mass has been given in Table \ref{T2} indicating higher
radius for higher temperatures.
Here, it should be noted that as it was seen from Fig. \ref{fig5}, by increasing the temperature, the equation of state of SQS becomes stiffer.
Now, we can conclude that in the finite temperature NJL model of SQS, the stiffer equation of state leads to the higher maximum gravitational mass for this compact object
(Table \ref{T2}).
This behavior has been also reported by Chu et al. \cite{Chu2017}.

%
\begin{table}[h]
\begin{center}
  \caption[]{Structure properties of SQS
at different temperatures.}\label{T2}
  \begin{tabular}{clclclclclclcl}
  \hline\noalign{\smallskip}
 $T (MeV)$ & $M_{max} (M_{\odot})$ & $R (km)$  &&  $\overline{\rm \rho} (10^{15} \frac{g}{cm^{3}})$ && $\sigma$ && $ Z_{s}$\\
 \hline\noalign{\smallskip}
 $30$ & 0.650 & 5.305 && 1.86 && 0.348  && 0.239 \\
 $45$ & 0.742 & 5.498 && 2.29 && 0.411 && 0.303 \\
 $70$ & 0.85 & 5.962 && 2.12 && 0.436 && 0.331  \\
 $90$ & 1.002 & 6.339 && 1.85 && 0.448 && 0.346  \\

  \noalign{\smallskip}\hline
  \end{tabular}
\end{center}
\end{table}
%
%
\subsection{Average density}
We can calculate the average density of the star using maximum mass ($M$) and radius ($R$) by,
\begin{equation}\label{17}
\overline{\rm \rho}=\frac{3M}{4 \pi R^{3}}.
\end{equation}
The results of this calculation are shown in Table \ref{T2}.  The minimum average density regarding Table \ref{T2}, $\overline{\rm \rho}=1.85 \times 10^{15} g/ cm^{3}$, is related to temperature of $90 MeV$  which is larger than the normal nuclear matter density, $\rho_{0}=2.7 \times 10^{14} g/ cm^{3}$.
Furthermore, the central density of SQS regarding Table \ref{T1} is about $7 \times 10^{15} g/ cm^{3}$, which is larger than average density of SQS at all temperatures.
\subsection{Compactness}
The compactness is a parameter to show the strength of gravity. It is calculated using the ratio of Schwarzschild radius to radius of star ($\sigma = R_{sch}/R$ where $R_{sch}=\frac{2GM}{c^{2}}$). As it is shown in Table
\ref{T2}, $\sigma$ is almost the same at all temperatures for SQS.
\subsection{Gravitational redshift}
\begin{figure}
\centering
\includegraphics[width=10cm, height=7cm]{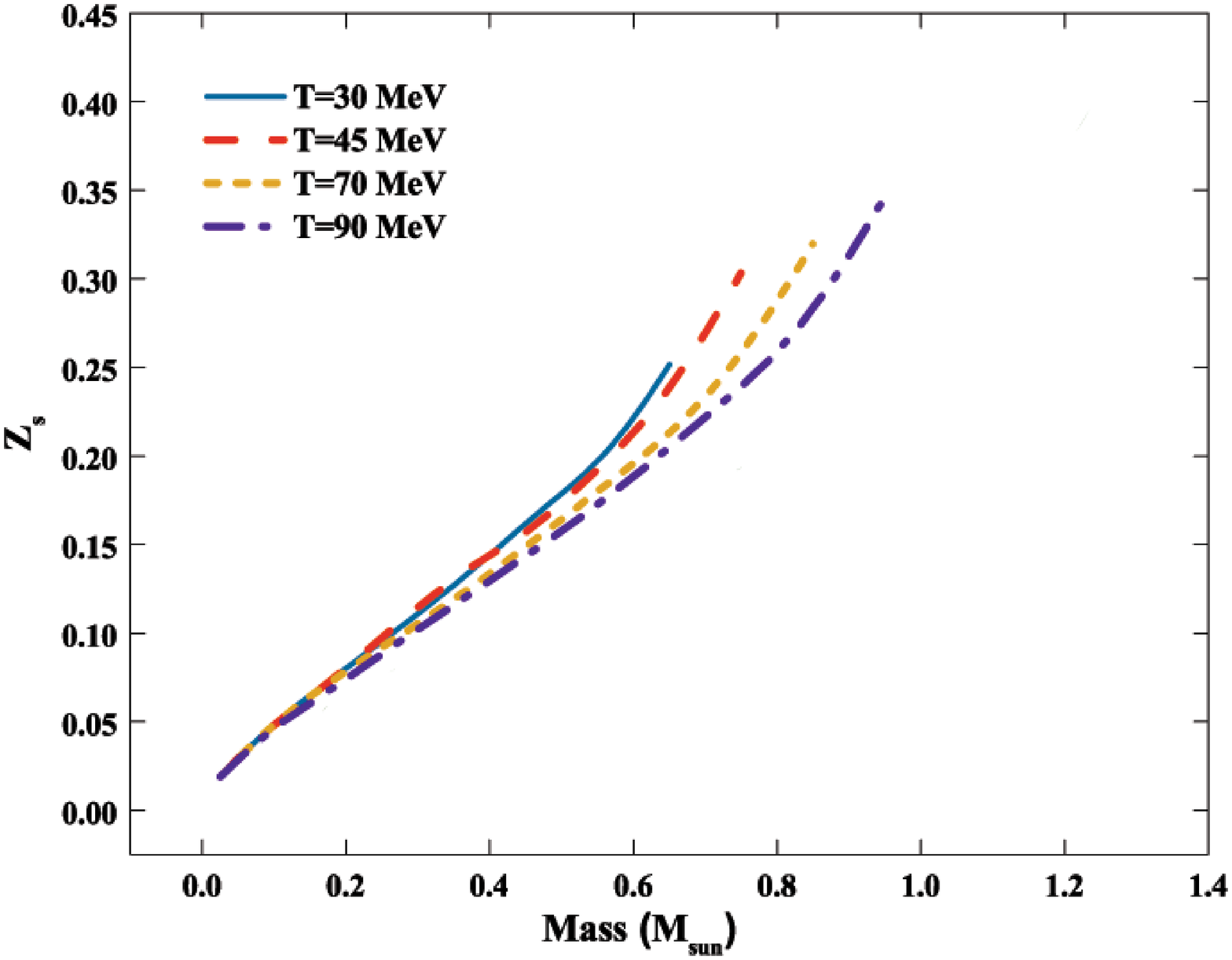}
\caption{The gravitational redshift of SQS as a function of the gravitational mass at different temperatures.} \label{fig8}
\end{figure}

The gravitational redshift is calculated as,
\begin{equation}\label{15}
Z_{s}=\frac{1}{\sqrt{1-\frac{2GM}{c^{2}R}}}-1,
\end{equation}
where $M$ is the maximum mass and $R$ is the radius of SQS.
We have plotted the gravitational redshift of SQS versus the gravitational mass at different temperatures in Fig. \ref{fig8}.
%
Obviously, it can be seen that for all temperatures, the gravitational redshift increases by increasing the gravitational mass to the value of the maximum limit.
Also, it is clear that the gravitational redshift increases by increasing temperature.
The results of gravitational redshift of SQS corresponding to maximum mass and temperature
 have been shown in the last column of Table \ref{T2}.
It can be seen that  the gravitational redshift increases as a function of maximum mass.
The maximum gravitational redshift is calculated, $z_{s}$=$0.346$, at temperature of $T=90$ $MeV$, that is about $59.34$$\%$ less than
critical value of gravitational redshift ($Z_{s}^{CL}$=$0.8509$) \cite{Haensel2007}. Furthermore, the gravitational redshift at temperature of $T$=$90$$MeV$ ($Z_{s}$=$0.346$) is about $0.4$$\%$
less than the observational result that is reported for quark star candidate $RX J185635-3754$ ($Z_{s}$=$0.35$ $\pm$ $0.15$) \cite{Prakash2003}.
\subsection{The mass of SQS in term of Planck mass}
In this section, we show that the mass of SQS can be expressed in term of the fundamental value of Planck mass, then we derive the relevant relation.
The repulsive nuclear force and the degeneracy pressure of fermions both are against gravity to avoid the collapse of compact stars. As we have mentioned, by phase transition of nucleons, the density of SQS is near and above the normal nuclear matter density. Therefore, using these facts, in the maximum value, we can consider the average density of SQS equal to the nuclear density, where the nuclear density is approximately defined as follow,
\begin{equation}\label{16}
\rho_{nuc}\simeq \frac{3m_{p}}{4\pi \lambda_{\pi}^{3}},
\end{equation}
where $m_{p}$ is the proton mass and $\lambda_{\pi}=\hbar/m_{\pi}c$ is the Compton wavelength of pion.
From previous sections, we use $R_{sch}$ and $\overline{\rm \rho}$ to derive the following equation,
\begin{equation}\label{17}
M\simeq (\frac{\hbar c}{G})^{3/2} \frac{1}{m_{p}^{2}} (\frac{\eta_{\pi}}{2\eta_{p}})^{3/2}
    \simeq M_{Ch} (\frac{\eta_{\pi}}{2\eta_{p}})^{3/2}
      \simeq m_{pl} \eta_{p}^{2} (\frac{\eta_{\pi}}{2\eta_{p}})^{3/2},
\end{equation}
where $M_{Ch}$ is the Chandrasekhar mass ($\simeq  (\frac{\hbar c}{G})^{3/2} \frac{1}{m_{p}^{2}}$), $m_{pl}$ is Planck mass,
$\eta_{p}=m_{pl}/m_{p}$ and $\eta_{\pi}=m_{pl}/m_{\pi}$ \cite{Burrows2014}.


\section{ The temperature dependence of graviational maximum mass of SQS}
In this section, we want to look at the behavior of the maximum gravitational mass of SQS which is calculated from different methods at finite temperature.

We have calculated the thermodynamic properties and structure of SQS  at finite temperature using MIT bag model with the fixed bag constant and density-dependent bag constant \cite{Bordbar2011vol54}.
It has been shown that the EOS of the system in both cases (fixed bag constant and density-dependent bag constant) becomes stiffer by increasing temperature.
Then, we have shown that the maximum gravitational mass and the corresponding radius decrease as a function of temperature in both mentioned cases.
For $B=90$$MeV$, the maximum gravitational mass and the corresponding radius  have been calculated to be $1.228$ $M_{\odot}$ and $7.073$ $km$ at $T=30$ $MeV$
and $1.04$ $M_{\odot}$ and $6.14$ $km$ at $T=80$ $MeV$. Also for density dependent bag constant the maximum gravitational mass has been changed
from $1.34$ to $1.12$ $M_{\odot}$ where temperature changed from $T=30$ $MeV$ to $T=80$ $MeV$. In the same way the radius of SQS decreases from $7.44$ to $6.57$ $km$.

We have also investigated the structure of spin-polarized strange quark star at finite temperature using MIT bag model with $B=90$$MeV$ \cite{Bordbar2011vol11},
and with a density-dependent bag constant \cite{Bordbar2012vol12}.
The EOS and the maximum gravitational mass and radius in Refs. \cite{Bordbar2013, Bordbar2014} were similar to the previous work \cite{Bordbar2011vol54}.
The maximum gravitational mass decrease from $1.171$ $M_{\odot}$ at $T=30$ $MeV$ to $1.16$ $M_{\odot}$ at $T=70$ $MeV$ using $B=90$ $MeV$, and the radius decreased from $7.27$ to $7.21$ $km$.
By considering a density dependent bag constant the maximum gravitational mass and the corresponding radius decreased from $1.15$$M_{\odot}$ and
$7.1$ $km$ at $T=30$ $MeV$ to $0.77$ $M_{\odot}$ and $6.89$ $km$ at $T=70$ $MeV$.

The structure of SQS has been calculated by Alaverdyan and Hajyan \cite{Alaverdyan2014}. They have considered the ultrarelativistic quarks in SQS and have calculated the EOS of the system using MIT bag model.
As they have reported, the EOS becomes stiffer as the temperature increases, where the radius of SQS versus temperature has been plotted.
It can be seen from this figure that the radius and the corresponding gravitational mass increase from $7.23$ $km$ and $0.49$ $M_{\odot}$ to $8.27$ $km$ and $0.77$ $M_{\odot}$
when temperature increases in a range from zero to $80$ $MeV$.

Compact strange stars with a medium dependence on gluons at finite temperature have been studied by Bagchi et al. \cite{Bagchi2006}.
Properties have been calculated using large color approximation with built-in chiral symmetry restoration in that paper.
Their calculations have shown that the stiffer EOS has been achieved at higher temperatures. Similarity, the maximum gravitational mass and the correspondig radius are larger at the lower temperatures.

As we have shown in the section \ref{secA} of the present paper, using NJL model creates a different behavior in gravitational mass as a function of temperature.
We can see from Table \ref{T2},
the maximum gravitational mass and radius increase by increasing temperature, although the EOS of the system becomes stiffer by increasing temperature.

A similar behavior with our current work also has been reported in Ref. \cite{Chu2017}. They have used NJL model as in our current paper.
They have plotted the equation of state at three different temperatures. It has been shown that EOS becomes stiffer by increasing temperature as we have shown in Fig. \ref{fig5}.
Furthermore, the gravitational mass as a function of temperature has been plotted and it has been reported that when temperature rises to $50$, $80$, and $100$ $MeV$, the maximum mass of quark stars
will reach $2.13$ $M_{\odot}$, $2.46$ $M_{\odot}$, and $2.71$ $M_{\odot}$, respectively.

In another work \cite{Dexheimer2013} the authors have compared the properties of the proto-quark star from different methods (Quark-mass density-dependent (QMDD) model, MIT bag model, and NJL model) at finite temperature.
They have reported the maximum gravitational mass which is calculated from QMDD and MIT bag models at different temperatures.
Their results show that investigation of the proto-quark star gives different behavior for mass-radius results as the function of temperature by MIT bag model.
Where they have considered the same conditions as in Ref. \cite{Bordbar2011vol54}, different behavior has been achieved for mass and radius in the temperature range. The gravitational mass increases
from $1.62$ to $1.65$ $M_{\odot}$ and the corresponding radius  increases from $9.01$ to $9.15$ $km$.
So, when the conditions are such as in Refs. \cite{Bordbar2013, Bordbar2014}, although the gravitational mass decreases from $2.02$ to $1.93$ $M_{\odot}$,
the radius increases from $9.04$ to $9.08$ $km$ by increasing temperature.
Using QMDD, they have considered two versions: $1)$ where the pressure at the density corresponding to the minimum of the free energy per baryon could be non-zero,
 depending on the matter studied (SM or 2QM) is noted as version 1 (QMDDv1) $2)$ presenting a remedy to the thermodynamical inconsistency, in such a way that the minimum of the energy per
baryon corresponds to the point of zero pressure, is noted as version 2 (QMDDv2). In both versions, they have considered different masses for strange quarks ($150$ and $100$ $MeV/c^{2}$ ).
Using QMDDv1, it has been shown that the maximum gravitational mass and the corresponding radius increase by increasing temperature for both strange quark masses
(from $2.28$ to $2.33$ $M_{\odot}$ and $12.05$ to $12.19$ $km$ for the first strange quark mass and from $2.26$ to $2.29$ $M_{\odot}$ and $11.75$ to $11.76$ $km$ for the second strange quark mass).
Using QMDDv2 when they have used strange quark mass equal to $150$ $MeV/c^{2}$ the maximum gravitational mass and corresponding radius increases by increasing temperature
(from $1.60$ to $1.62$ $M_{\odot}$ and $8.42$ to $8.46$ $km$),
but when they have considered the strange quark mass equal to $100$ $MeV/c^{2}$ there is an inverse behavior for the maximum gravitational mass and radius in the temperature range
 (the gravitational mass decrease from $1.59$ to $1.58 $ $M_{\odot}$ and the radius decrease from $8.22$ to $8.16$ $km$).


\section{Summary and conclusions}
In this paper, we have calculated the thermodynamic properties of the strange quark matter (SQM)
at finite temperatures using NJL model, and we have investigated the structure of strange quark stars (SQS).
We have calculated free energy and equation of state (EOS) of SQM by considering the dynamical mass.
We have shown that free energy increases the corresponding baryonic density. In addition, free energy decreases by increasing temperature at a specified density.
Also, our results indicate that pressure increases proportional to the density and EOS of SQM becomes stiffer as a function of temperature.
Furthermore, we have investigated the  energy conditions and stability of EOS. We have shown that EOS of our system satisfies both energy conditions and stability.

Later, we studied the structure of SQS using the general relativistic TOV equations and boundary conditions. We calculated the maximum gravitational mass and the corresponding
radius of SQS at different temperatures.
Follow up on the structure of the star, we have calculated other parameters such as the Schwarzschild Radius, average density of SQS,
compactness and gravitational redshift.
We have shown that the gravitational mass and radius of SQS rapidly increases by increasing temperature and
we have compared the behavior of temperature dependent maximum gravitational mass for different methods.
We have shown that the average density of SQS is more than the normal nuclear matter density.
In addition, our calculations show that the compactness of SQS is almost the same at all temperatures.
We have also investigated the gravitational redshift ($Z_{s}$) of SQS at different temperatures and found that $Z_{s}$
 increases by increasing temperature. Comparison with the observational results clarify that the gravitational redshift of SQS at
temperature of $90$$MeV$ is just $\%$$3$ less than the gravitational redshift of
$RX J185635-3754$.
Finally we have derived the relation between mass of SQS and Plank mass.

\section*{Acknowledgements}
{We wish to thank Shiraz University
Research Council. This work has been supported by Research Institute for Astronomy
and Astrophysics of Maragha. 
G. H. Bordbar wishes to thank A. Broderick (University of Waterloo) for his useful comments and discussions during this work.
G. H. Bordbar also wishes to thank Physics Department of University of Waterloo
for the great hospitality during his sabbatical.}


\end{document}